\documentclass[12pt]{article}
\usepackage{latexsym}
\usepackage{amsmath}
\usepackage{amsfonts}
\usepackage{amssymb}

\title{N-Galilean conformal algebras and higher derivatives Lagrangians   }
\author{K. Andrzejewski\thanks{e-mail: k-andrzejewski@uni.lodz.pl},
J. Gonera\\
\small Department of Theoretical Physics and Computer Science, \\
\small University of \L\'od\'z,\\
\small Pomorska 149/153, 90-236 {\L}\'od\'z, Poland
}
\date{}
\begin{document}
\maketitle 
\begin{abstract}
It is shown that the N-Galilean conformal algebra, with N-odd, is the maximal symmetry algebra of the free Lagrangian involving $\frac{N+1}{2}$-th  order time derivative.
\end{abstract}
\section{Introduction}
The structure of conformal transformations in nonrelativistic space-time is more involved than in the relativistic case. In fact,  there exists the infinite family of conformal  extensions of Galilei algebra (or group)  indexed by the integer $N$ \cite{b1}-\cite{b3}which are called $N$-Galilean conformal algebras. This family naturally decomposes into two subfamilies depending on whether $N$ is odd or even. In the former case the $N$-Galilean conformal algebra admits  central extension while in the latter it doesn't \cite{b4}-\cite{b6} (except in two-dimensional  space  where the central extension is allowed for any $N$).
\par
The $N=1$ member of conformal family is the famous Schr\"odinger algebra (discovered already in nineteenth century \cite{b7}) \cite{b1,b2},\cite{b8}-\cite{b15}. It has a clear dynamical interpretation as  the maximal symmetry algebra/group of free Schr\"odinger equation \cite{b10} (and, also, the free classical Lagrangian). The natural question arises concerning the dynamical realizations of other members of nonrelativistic conformal family. Such realizations has been constructed and involve higher-derivative Lagrangians  \cite{b6},\cite{b16}-\cite{b19}. In fact, it has been shown \cite{b19} (see also \cite{b6})  with the help of orbit method \cite{b20} that for the conformal algebras which admit central extension (i.e. $N$ odd for $d\geq 3$ and  $N$ arbitrary for $d=2$) there exists  dynamical realization such that the time development of "external" variables is described by Ostrogradski  Hamiltonian \cite{b21} corresponding to the  free Lagrangian involving $\frac{N+1}{2}$-th order time derivatives. It can be also shown that the action of conformal group on phase space translates to the action on coordinate space as Noether symmetry \cite{b22}.
\par 
The case of general $N$ (i.e. including the conformal algebras which do not admit central extension) is more involved. The dynamical systems admitting  $N$-Galilean conformal algebra as symmetry algebra  can be constructed \cite{b23}-\cite{b24} using the method of nonlinear realizations \cite{b25}. However, the problem is that the existence of Lagrangian and Hamiltonian formalism cannot be in this case taken for granted. 
\par
Coming back to the case of $d=3$, $N$-odd algebra the question arises whether they can be characterized in the same manner as Schr\"odinger algebra 
%%%
i.e. as maximal symmetry algebras of free motion described by higher-derivative Lagrangians or,on the quantum level,as  maximal symmetry algebras of the Schr\"odinger equation associated  the with the Ostrogradski Hamiltonian of this Lagrangian. In the present note we answer to  the first  part of this question  giving a simple proof of the fact $N$-Galilean conformal algebra/group is the maximal symmetry algebra/group of free Lagrangian involving $\frac {N+1}{2}$ order derivatives; the problem of the invariance of the relevant Schr\"odinger equation will be discussed elsewhere \cite{b26}.
\par
The higher-derivative theory is defined by the Lagrangian 
\begin{equation}
\label{e1}
L=L(\vec q,\dot {\vec q},\ldots,{\vec {q}}^{(n)}).
\end{equation}
The dynamics is derived from the least action principle 
\begin{equation}
\label{e2}
\delta S=0,
\end{equation}
where 
\begin{equation}
\label{e3}
S=\int _{t_0}^{t_1}dt L(\vec q,\dot{\vec q},\ldots, \vec q^{(n)}),
\end{equation}
and the variation  (\ref{e2}) is performed under the boundary conditions $\delta \vec q(t)=\delta \dot{\vec q}(t)=\ldots=\delta \vec q^{(n-1) }(t)=0$ for $t=t_0,t_1$. 
 The resulting equation of motion reads 
 \begin{equation}
\label{e4}
\sum_{k=0}^{n}(-1)^k\frac{d^k}{dt^k}\left(\frac{\partial L }{\partial \vec q^{(k)}}\right)=0.
\end{equation}
The free motion is defined by the following choice of the Lagrangian 
\begin{equation}
\label{e5}
L=\frac{m}{2}\left (\frac{d^n\vec q}{dt^n}\right)^2,
\end{equation}
where $\vec q$ is the  coordinate in three-dimensional  (in general $d$-dimensional) Euclidean space and $m$  is a "mass" parameter of dimension $\textrm{kg}\cdot\textrm{s}^{2(n-1)}.$\par 
The  dynamical equation (\ref {e4}) takes the form 
\begin{equation}
\label{e6}
\frac{d^{2n}\vec q}{dt^{2n}}=0.
\end{equation}
Analogously to the standard case we define the Noether symmetry as the transformation 
\begin{equation}
\label{e7}
t'=t'(t), \quad {\vec{q'} }(t')=\vec{ q'}(\vec q,t),
\end{equation}
obeying
\begin{equation}
\label{e8}
L(\vec{ q'},\dot {\vec {q'}},\ldots,{\vec {q'}}^{(n)})\frac{dt'}{dt}=L(\vec q,\dot {\vec q},\ldots,{\vec {q}}^{(n)})+\frac{df(\vec q,\dot {\vec q},\ldots,{\vec {q}}^{(n-1)})}{dt}.
\end{equation}
\par 
We are looking for the maximal continuous (i.e. Lie) group of transformations (\ref{e7}) which are Noether symmetries of the Lagrangian (\ref {e5}). To this end consider the infinite form of eqs. (\ref {e7})
\begin{equation}
\label{e9}
t'=t+\epsilon\psi(t) , \quad {\vec{q'} }(t')=\vec{ q}+\epsilon\vec \phi(\vec q,t).
\end{equation}
Noting that $\frac{dt'}{dt}=1+\epsilon\dot\psi$ 
and 
\begin{align}
\label{e10}
\frac{d^n\vec {q'}}{dt'^n}&\cong \left((1-\epsilon\dot \psi)\frac{d}{dt}\right)^n(\vec q+\epsilon\vec\phi)\nonumber \cong\\
&\frac{d^n\vec q}{d t^n}+\epsilon\frac{d^n\vec \phi}{dt^n}-\epsilon\sum_{k=0}^{n-1}\sum_{l=0}^{k}\dbinom{k}{l}\psi^{(l+1)}\frac{d^{n-l}\vec q}{dt^{n-l}},
\end{align}
one finds
\begin{align}
\label{e11}
\frac{m}{2}\left(\frac{d^n\vec{q'}}{dt'^n}\right)^2\frac{dt'}{dt}=&
\frac{m}{2}\left(\frac{d^n\vec{q}}{dt^n}\right)^2 +\epsilon\frac{m}{2}\left[\dot\psi\frac{d^n\vec q}{dt^n}\right. +\nonumber\\
&2\left.\left(\frac{d^n\vec \phi}{dt^n}-\sum_{k=0}^{n-1}\sum_{l=0}^{k}\dbinom{k}{l}\psi^{(l+1)}\frac{d^{n-l}\vec q}{dt^{n-l}}\right)\right]\frac{d^n\vec q}{dt^n}.
\end{align}
Therefore, by replacing $f \rightarrow \epsilon f$ in eq. (\ref {e8}) we obtain 
\begin{align}
\label{e12}
\frac{m}{2}\left[\dot\psi\frac{d^n\vec q}{dt^n}\right. +&
2\left.\left(\frac{d^n\vec \phi}{dt^n}-\sum_{k=0}^{n-1}\sum_{l=0}^{k}\dbinom{k}{l}\psi^{(l+1)}\frac{d^{n-l}\vec q}{dt^{n-l}}\right)\right]\frac{d^n\vec q}{dt^n}=  \nonumber \\ 
&\sum_{k=0}^{n-1}\frac{\partial f}{\partial {\vec q}^{(k)}}+\frac{\partial f}{\partial t}.
\end{align}
The right-hand side  of eq. (\ref{e12}) is linear in $\frac{d^n\vec q}{dt^n}$. The coefficient in front of $\left(\frac{d^n\vec q}{dt^n}\right)^2$ on the left-hand side must therefore vanish. This yields 
\begin{equation}
\label{e13}
(1-2n)\dot\psi\delta_{ab}+\left(\frac{\partial\phi_a}{\partial q_b}+\frac{\partial\phi_b}{\partial q_a}\right)=0.
\end{equation}
Differentiating eq. (\ref{e13}) with respect to $q_c$ and combining these equations obtained by cyclic permutations of $a,b,c$ we arrive at $\frac{\partial^2\phi_a}{\partial q_b\partial q_c}=0$  and consequently 
\begin{equation}
\label{e14}
\frac{\partial\phi_a}{\partial q_b}=\left(\frac{2n-1}{2}\right)\dot\psi\delta_{ab}+\omega_{ab}(t),\quad \omega_{ab}=-\omega_{ba};
\end{equation}
\begin{equation}
\label{e15}
\phi_a=\left(\frac{2n-1}{2}\right)\dot\psi q_{a}+\omega_{ab}(t)q_b+\chi_a(t).
\end{equation}
Once eq. (\ref{e15}) is satisfied, the left hand side of (\ref{e12}) is linear in $\frac{d^n\vec q}{st^n}$ so that 
\begin{equation}
\label{e16}
f=f(\vec q^{(n-1)}).
\end{equation}
As a result no $\vec q^{(k)}$, $0\leq k\leq n-2$, can appear on the left-hand  side. Together with the integrability condition 
$\frac{\partial^2 f}{\partial q^{(n-1)}_a\partial q_b^{(n-1)}}$ this implies
\begin{equation}
\label{e17}
\chi_a^{(n+1)}=0, \quad \dot\omega_{ab}=0, \quad \psi^{(3)}=0, 
\end{equation}
and
\begin{equation}
\label{e18}
\psi=\tau+\lambda t+ct^2,\quad \chi_a=\sum_{k=0}^nv_{ak}t^k,
\end{equation}
or
\begin{equation}
\label{e19}
\begin{split}
&\delta t=\tau+\lambda t+ct^2,\\
&\delta q_a=\left(\frac{2n-1}{2}\right)(\lambda +2ct)q_a+\omega_{ab}q_b+\sum_{k=0}^nv_{ak}t^k.
\end{split}
\end{equation}
By considering the terms containing particular parameters we can identify the differential realizations of the symmetry algebra:
\begin{equation}
\begin{split}
&H=i\frac{\partial}{\partial t},\quad D=-i\frac{\partial}{\partial t}-i\left(\frac{2n-1}{2}\right)\vec q\frac{\partial}{\partial\vec q},\\
&K=it^2\frac{\partial}{\partial t} +i\left(2n-1\right)t \vec q\frac{\partial}{\partial\vec q},\\
&\vec J=-i\vec q\times \frac{\partial}{\partial \vec q},\quad C_{ak}=i(-1)^kt^k\frac{\partial }{\partial q_a}.
\end{split}
\end{equation}
They obey $N$-Galilean conformal algebra with $N=2n-1$ $(a,b,c=1,2,3; k=1,\ldots,n)$:
\begin{equation}
\begin{split}
&[D,H]=iH,\quad [D,K]=-iK,\quad [K,H]=2iD,\\
&[J_a,C_{bk}]=i\epsilon_{abc}C_{ck},\\
&[H,C_{ak}]=-ikC_{ak-1},\\
&[D,C_{ak}]=i\left(\frac{2n-1}{2}-k\right)C_{ak},\\
&[K,C_{ak}]=i(2n-1-k)C_{ak+1}.\\
\end{split}
\end{equation}
The algebra obtained is not centrally extended in spite of the fact that it does admit such an extension. The central charge appears only on the Hamiltonian level. 
\par We have shown that the $N$-Galilean conformal algebras, with $N$ odd, can be interpreted as maximal symmetry algebras of free motion described by higher-derivative Lagrangians. This is the generalization  of Niederer's characterization of Schr\"odinger algebra as the maximal symmetry algebra of standard free dynamics \cite{b10}.
%%%%%%%%%%%%%% 
\par
{\bf Acknowledgments} The authors would like to thank Professor Piotr Kosi\'nski for helpful discussions and  useful remarks.
This work  is supported  in part by  MNiSzW grant No. N202331139.

\end{document}